# Delayed choice relativistic quantum bit commitment with arbitrarily long commitment time

*Muhammad Nadeem*
*Department of Basic Sciences,*
*School of Electrical Engineering and Computer Science*
*National University of Sciences and Technology (NUST)*
*H-12 Islamabad, Pakistan*
*muhammad.nadeem@seecs.edu.pk*

We propose here a two-round relativistic bit commitment scheme where committer commits in the first round and then confirms his/her commitment in the second round. The scheme offers indefinite commitment time where both committer and receiver extract non-locally correlated measurement outcomes during the scheme that can be stored and revealed after arbitrarily long time. We show that the proposed scheme turns out to be a multiparty bit commitment scheme where both parties commit and reveal simultaneously. The multiparty generalization would have applications in business and secure multiparty computations such as managing joint bank account, holding joint shares in stock exchange and blind bidding. The same bit commitment scheme can also be used for the commitment of arbitrarily long classical bit strings. The scheme can be applied efficiently with existing technologies; entanglement is required only for time $t=x/c$ where $x$ is the special distance, can be as small as possible, between the committer and receiver.

**Introduction**
A bit commitment is an important cryptographic task between two mistrustful parties, a committer and a receiver. In general, committer commits to a specific bit by giving some information to the receiver and then unveils his/her commitment at some later time. Bit commitment is said to be unconditionally secure if it fulfils following security requirements: (i) receiver should not be able to extract the committed bit value during the scheme. (ii) When committer reveals, it must be possible for receiver to know the genuine bit value with absolute guarantee. (iii) Committer should not be able to reveal a bit different from the committed one.

    In non-relativistic classical cryptography, secure bit commitment scheme based on unproven computational hardness is impossible. Similarly, non-relativistic quantum bit commitment schemes[1-3] are also insecure against Mayers and Lo-Chau quantum attacks[4-7] if committer and receiver do not pre-share any data.

    However, in relativistic cryptographic settings, both classical[8,9] and quantum bit commitment protocols[10,11] have been presented and proved to be secure against general classical and quantum attacks[12,13]. Moreover, relativistic quantum bit commitment protocols have been experimentally demonstrated, using quantum communication and special relativity, with commitment times of 15ms[14] and 30µs[15], respectively. The commitment time is equivalent to the communication time between committer and one of the receiver's agents. On the other hand, relativistic classical bit commitment protocol[8] allows committer to commit for arbitrary duration but it requires multiple rounds with exponential increase in communication which makes it impractical. Recently, Lunghi *et al* proposed a relativistic classical multi-round bit commitment scheme[16] with commitment time of 212ms, longer than communication time between the locations of the agents.

Recently, we proposed that combination of quantum non-locality and causality allows relativistic quantum cryptography to define many interesting mistrustful cryptographic tasks[17-20] that were previously considered to be impossible[21-24]. The general framework for relativistic quantum cryptography[16] directly leads to (i) a new notion of OT where both the data transferred and the transfer position remain oblivious, (ii) deterministic two-sided TPSC, (iii) asynchronous ideal coin tossing with zero bias, and (iv) Unconditionally secure bit commitment with maximum commitment time of $2x/c$ while minimum commitment time of $x/c$ where $x$ is the special distance between committer and receiver.

These possibilities of wide range of cryptographic tasks in relativistic quantum cryptography show that relativity adds its weight, and hence gives more power, towards quantum cryptography. It motivates us to define more systematic and widely applicable bit commitment scheme based on relativistic quantum principles. For example, world is no longer divided between Alice and Bob but now we have a more complex, multi-party setting. The communication between these distant parties takes some finite time. So it is natural to ask how many parties can commit and reveal simultaneously. Another question would be to ask how long a commitment can sustain securely. More importantly, are multiple rounds of communication necessary to obtain arbitrarily long commitment time? In this article, we tried to address these questions and find very interesting answers.

We propose here a delayed choice relativistic quantum commitment scheme where committer commits to a specific bit by sharing an EPR pair $|ij\rangle$ with receiver in the first round. At some later time that comes from the scheme, committer applies unitary transformations $\sigma_z^i \sigma_x^j$ on his/her retained qubit and confirms her commitment by sending retained qubit to the receiver in second round. The proposed relativistic quantum scheme leads to (i) a secure commitment scheme between a committer and receiver. (ii) Secure commitment scheme between two parties where both commit and reveal simultaneously. The scheme validates commitments if both parties are fair otherwise aborts if any party tries to cheat. (iii) Commitment scheme for the commitment of arbitrarily long classical bit strings. (iv) Arbitrarily long commitment time. The multiparty bit commitment scheme would have applications in business such as managing joint bank account, holding joint shares in stock exchange and blind bidding. More generally, it can implement multiparty secure computation between distant parties.

Interestingly, we found that arbitrarily long commitment time can be achieved securely by using only two-rounds of communication. Moreover committer and receiver do not need of quantum memory for storing quantum data for longer times; both committer and receiver extract non-locally correlated measurement outcomes in the commitment phase that can be stored and revealed after arbitrarily long time. Our scheme is purely relativistic quantum mechanical; it does not require any classical communication between committer and receiver.

**Space time geometry and security criteria**
Suppose committer, receiver and their agents have secure point like laboratories at priorly announced positions in Minkowski space time. Both parties can securely communicate quantum information with their respective agents by sending signals at speed of light c. The time for information processing at their laboratories is assumed to be negligibly small as compared to communication time between them.

The proposed scheme attains unconditional security through combination of quantum non-local correlations generated by entanglement swapping[25,26] and then teleportation[27] over swapped entangled quantum systems. We denote four Bell states as $\zeta^+ = |00\rangle, \eta^+ = |01\rangle, \zeta^- = |10\rangle$ and $\eta^- = |11\rangle$ while qubits of these Bell pairs as $\{\alpha, \alpha_0\}$ in case of Alice and $\{\beta, \beta_0\}$ if Bob prepares any one of these Bell states.

$$\zeta^+ = |00\rangle = \frac{|00\rangle + |11\rangle}{\sqrt{2}} \tag{1.1}$$

$$\eta^+ = |01\rangle = \frac{|01\rangle + |10\rangle}{\sqrt{2}} \tag{1.2}$$

$$\zeta^- = |10\rangle = \frac{|00\rangle - |11\rangle}{\sqrt{2}} \tag{1.3}$$

$$\eta^- = |11\rangle = \frac{|01\rangle - |10\rangle}{\sqrt{2}} \tag{1.4}$$

**Security definition-I:** Suppose two parties A and B share their secretly prepared EPR pairs $|\alpha\alpha_0\rangle$ and $|\beta\beta_0\rangle$ with a third party C respectively. If C applies Bell state measurement (BSM) on $|\alpha_0\rangle$ and $|\beta_0\rangle$, A and B gets entangled in one of the Bell states $|\alpha\beta\rangle$ whose exact identity remains unknown to all three parties unless they communicate and gather their secret information at one point in their causal future.

Here Bell state $|\alpha\alpha_0\rangle$ is known only to Alice while $|\beta\beta_0\rangle$ to Bob. By performing measurement of Bell operator[26] on qubits $|\alpha_0\rangle$ and $|\beta_0\rangle$, C projects initial system $|\alpha\alpha_0\rangle \otimes |\beta\beta_0\rangle$ into $|\alpha_0\beta_0\rangle \otimes |\alpha\beta\rangle$ where Bell state $|\alpha_0\beta_0\rangle$ is known only to C. If $|\alpha\alpha_0\rangle \otimes |\beta\beta_0\rangle = \zeta^\pm \otimes \zeta^\pm$ or $|\alpha\alpha_0\rangle \otimes |\beta\beta_0\rangle = \eta^\pm \otimes \eta^\pm$ then

$$|\alpha\alpha_0\rangle \otimes |\beta\beta_0\rangle \rightarrow |\alpha_0\beta_0\rangle \otimes |\alpha\beta\rangle = \frac{1}{2}(\zeta^+ \otimes \zeta^+ + \eta^+ \otimes \eta^+ + \zeta^- \otimes \zeta^- + \eta^- \otimes \eta^-) \tag{2.1}$$

If $|\alpha\alpha_0\rangle \otimes |\beta\beta_0\rangle = \zeta^\pm \otimes \eta^\pm$ or $|\alpha\alpha_0\rangle \otimes |\beta\beta_0\rangle = \eta^\pm \otimes \zeta^\pm$ then

$$|\alpha\alpha_0\rangle \otimes |\beta\beta_0\rangle \rightarrow |\alpha_0\beta_0\rangle \otimes |\alpha\beta\rangle = \frac{1}{2}(\zeta^+ \otimes \eta^+ + \eta^+ \otimes \zeta^+ + \zeta^- \otimes \eta^- + \eta^- \otimes \zeta^-) \tag{2.2}$$

If $|\alpha\alpha_0\rangle \otimes |\beta\beta_0\rangle = \zeta^\pm \otimes \zeta^\mp$ or $|\alpha\alpha_0\rangle \otimes |\beta\beta_0\rangle = \eta^\pm \otimes \eta^\mp$ then

$$|\alpha\alpha_0\rangle \otimes |\beta\beta_0\rangle \rightarrow |\alpha_0\beta_0\rangle \otimes |\alpha\beta\rangle = \frac{1}{2}(\zeta^+ \otimes \zeta^- + \eta^+ \otimes \eta^- + \zeta^- \otimes \zeta^+ + \eta^- \otimes \eta^+) \tag{2.3}$$

If $|\alpha\alpha_0\rangle \otimes |\beta\beta_0\rangle = \zeta^\pm \otimes \eta^\mp$ or $|\alpha\alpha_0\rangle \otimes |\beta\beta_0\rangle = \eta^\pm \otimes \zeta^\mp$ then

$$|\alpha\alpha_0\rangle \otimes |\beta\beta_0\rangle \rightarrow |\alpha_0\beta_0\rangle \otimes |\alpha\beta\rangle = \frac{1}{2}(\zeta^+ \otimes \eta^- + \eta^+ \otimes \zeta^- + \zeta^- \otimes \eta^+ + \eta^- \otimes \zeta^+) \tag{2.4}$$

Equations (2.1)-(2.4) show that regardless of the initially shared Bell states $|\alpha\alpha_0\rangle$ and $|\beta\beta_0\rangle$, four Bell state measurement outcomes $|\alpha_0\beta_0\rangle \in \{\zeta^+, \eta^+, \zeta^-, \eta^-\}$ on C side are equally likely, each with probability of ¼. Hence, the exact identity of $|\alpha\beta\rangle$ can only be known to someone who knows all three Bell states $|\alpha\alpha_0\rangle$, $|\beta\beta_0\rangle$ and $|\alpha_0\beta_0\rangle$. Security definition-I holds true even if A and C or B and C collaborate.

**Security definition-II:** Suppose two parties A and B share an EPR pair $|\alpha\beta\rangle$ whose exact identity is unknown to both A and B. If A teleports a quantum state $|\varphi\rangle$ prepared in publically known orthogonal basis to B by performing measurement of Bell operator[26] on state $|\varphi\rangle$ and entangled qubit $|\alpha\rangle$, he gets one of the Bell states $\{\zeta^+, \eta^+, \zeta^-, \eta^-\}$ while B's entangled half becomes $|\psi\rangle = \sigma_i |\varphi\rangle$. The Pauli encoding $\sigma_i \in \{I, \sigma_x, \sigma_z, \sigma_z\sigma_x\}$ remains unknown to both A and B unless they communicate and gather information $|\psi\rangle$, $|\varphi\rangle$ and exact Bell state measurement outcome $\{\zeta^+, \eta^+, \zeta^-, \eta^-\}$ at one point in their causal future. If $|\alpha\beta\rangle = \zeta^+$ then

$$|\varphi\rangle \otimes |\alpha\beta\rangle = \frac{1}{2}(\zeta^+ \otimes |\varphi\rangle + \eta^+ \otimes \sigma_x|\varphi\rangle + \zeta^- \otimes \sigma_z|\varphi\rangle + \eta^- \otimes \sigma_x\sigma_z|\varphi\rangle) \quad (3.1)$$

If $|\alpha\beta\rangle = \eta^+$ then

$$|\varphi\rangle \otimes |\alpha\beta\rangle = \frac{1}{2}(\zeta^+ \otimes \sigma_x|\varphi\rangle + \eta^+ \otimes |\varphi\rangle + \zeta^- \otimes \sigma_x\sigma_z|\varphi\rangle + \eta^- \otimes \sigma_z|\varphi\rangle) \quad (3.2)$$

If $|\alpha\beta\rangle = \zeta^-$ then

$$|\varphi\rangle \otimes |\alpha\beta\rangle = \frac{1}{2}(\zeta^+ \otimes \sigma_z|\varphi\rangle + \eta^+ \otimes \sigma_x\sigma_z|\varphi\rangle + \zeta^- \otimes |\varphi\rangle + \eta^- \otimes \sigma_x|\varphi\rangle) \quad (3.3)$$

If $|\alpha\beta\rangle = \eta^-$ then

$$|\varphi\rangle \otimes |\alpha\beta\rangle = \frac{1}{2}(\zeta^+ \otimes \sigma_x\sigma_z|\varphi\rangle + \eta^+ \otimes \sigma_z|\varphi\rangle + \zeta^- \otimes \sigma_x|\varphi\rangle + \eta^- \otimes |\varphi\rangle) \quad (3.4)$$

It shows that regardless of the initially shared Bell state $|\alpha\beta\rangle$, four Bell state measurement outcomes $\{\zeta^+, \eta^+, \zeta^-, \eta^-\}$ on Alice side and hence four possibilities of Pauli encoding $\sigma_i$ on Bob's entangled half are equally likely, each with probability of ¼. Hence, the exact identity of Pauli encodings $\sigma_i$ can only be known to someone who knows both quantum information $|\psi\rangle$, $|\varphi\rangle$ and one of the foul possible Bell state measurement outcome $\{\zeta^+, \eta^+, \zeta^-, \eta^-\}$. Security definition-II holds true even if $|\varphi\rangle$ is known to both A and B.

**Delayed choice Relativistic quantum bit commitment scheme**
In the proposed delayed choice relativistic quantum bit commitment scheme committer commits to a specific bit by sharing an EPR pair $|ij\rangle$ with receiver in the first round. At some later time that comes from the scheme, committer applies unitary transformations $\sigma_z^i \sigma_x^j$ on his/her retained qubit and confirms her commitment by sending retained qubit to the receiver in second round. The relativistic quantum bit commitment scheme can be used to imply single-party commitment scheme and multiparty party commitment scheme with arbitrarily long commitment time and choice of committed data; single scheme works for the commitment of both single bit and commitment of arbitrarily long classical bit stings.
**Code:** Let's suppose measurement basis {0,1} are known to both Alice and Bob (priorly decided somewhere in their causal past) and they use the following code: Bell states $\zeta \in \{\zeta^+, \zeta^-\}$ correspond to classical 2-bit string $d_1 d_2 \in \{00, 10\}$ (or classical bit $d_c = d_2 = 0$) while those of $\eta \in \{\eta^+, \eta^-\}$ correspond to classical 2-bit string $d_1 d_2 \in \{01, 11\}$ (or classical bit $d_c = d_2 = 1$). That is, committer commits herself to the bit value $d_c = 0$ or $d_c = 1$ by preparing EPR pair $\zeta \in \{\zeta^+, \zeta^-\}$ or $\eta \in \{\eta^+, \eta^-\}$.

**Single-party commitment scheme:** Suppose receiver Bob, his agent $B_0$ and committer Alice are space like separated and occupy secure locations $(0,0)$, $(x,0)$ and $(2x,0)$ in Minkowski space time respectively. Suppose Bob and his agent $B_0$ secretly agree on a Bell pair $|\beta \beta_0\rangle$ and quantum state $|\varphi\rangle \in \{0,1\}$ somewhere in their causal past unknown to committer Alice. Explicit procedure for bit commitment scheme between committer Alice and receiver Bob is described below:

**Phase-I, t=0:** Both committer Alice and receiver Bob share Bell pairs $|\alpha \alpha_0\rangle \in \{\zeta, \eta\}$ and $|\beta \beta_0\rangle \in \{\zeta, \eta\}$ with $B_0$ respectively. By doing this, Alice commits herself to a 2-bit string $\alpha \alpha_0$.

**Phase-II, t=x/c:** $B_0$ performs BSM on qubits $|\alpha_0\rangle$ and $|\beta_0\rangle$ in his possession and stores classical BSM result $b_0 b_0'$. This measurement of $B_0$ projects the qubits in possession of Alice and Bob into one of the Bell states $|\alpha \beta\rangle \in \{\zeta^+, \eta^+, \zeta^-, \eta^-\}$. Simultaneously, Bob teleports a quantum state $|\varphi\rangle \in \{0,1\}$ to Alice and sends his BSM result $bb'$ to $B_0$ securely. As a result, Alice's half $|\alpha\rangle$ becomes one of the corresponding four possible states $|\psi\rangle = \sigma_i |\varphi\rangle$ where $\sigma_i$ is teleportation encoding. Instantly, Alice applies unitary operator $\sigma_z^\alpha \sigma_x^{\alpha_0}$ on her qubit $|\alpha\rangle$ and confirms her commitment by sending $|\psi'\rangle = \sigma_z^\alpha \sigma_x^{\alpha_0} \sigma_i |\varphi\rangle$ to $B_0$.

**Phase-III, t=2x/c:** $B_0$ measures $|\psi'\rangle$ in the $\{0,1\}$ basis and stores result $\psi'$.

**Phase-IV, t=T:** Alice reveals her commitment by announcing identity of $\alpha \alpha_0$ at some time t=T of her choice. Now $B_0$ can find the exact identity of swapped entangled state $|\alpha \beta\rangle$, and (hence) the exact teleportation encoding $\sigma_i$. If Alice's announcement is consistent with non-locally correlated Pauli encoding $\sigma_i$, $B_0$ validates Alice commitment otherwise aborts.

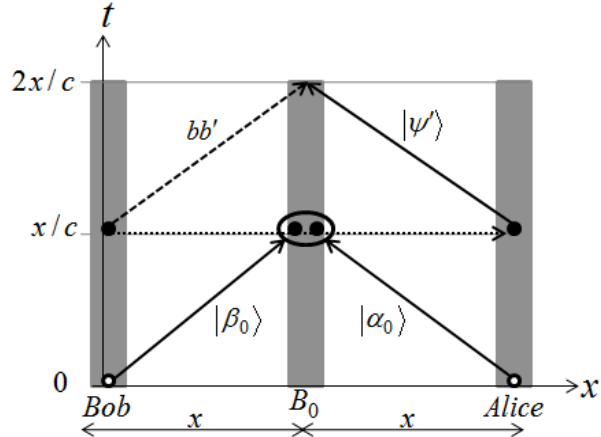

**Figure 1:** Relativistic quantum bit commitment scheme between committer Alice and receiver Bob and his agent $B_0$. At priorly agreed time t=0, Alice makes a commitment by sharing $|\alpha_0\rangle$ with $B_0$ and confirms her commitment by sending $|\psi'\rangle = \sigma_z^\alpha \sigma_x^{\alpha_0} \sigma_i |\varphi\rangle$ to $B_0$ at some later time t=x/c. Solid arrows represent flying qubits, dotted arrow shows teleportation while dashed arrow represents classical information secured by quantum channel between Bob and his agent.

We would like to highlight here that even though Alice shares EPR pair $|\alpha\alpha_0\rangle$ with $B_0$ at time t=0 to make initial commitment, her commitment gets mature at time t=$x$/c when $B_0$ receives $|\alpha_0\rangle$, Bob teleports $|\varphi\rangle$ to Alice, and Alice applies $\sigma_z^\alpha \sigma_x^{\alpha_0}$ on her qubit $|\alpha\rangle$ and returns to Bob. That is, Alice have a choice and can change her commitment (EPR pair) during the time laps from t=0 to t=$x$/c by applying specific Pauli transformations on retained qubit $|\alpha\rangle$ and it cannot be considered as successful cheating. However, after time t=$x$/c when she returns $|\psi'\rangle = \sigma_z^\alpha \sigma_x^{\alpha_0} |\psi\rangle$ to $B_0$ and confirms her commitment, it is impossible for Alice to alter her commitment.

**Multiparty-commitment scheme:** Suppose Alice and Bob are space like separated and occupy secure locations (0,0) and (2$x$,0) in Minkowski space time respectively. In the commitment phase, both Alice and Bob commit simultaneously by sharing an EPR pair with authority C at position ($x$,0). Here authority C is assumed to be a joint center (stock exchange) where both Alice and Bob have their representatives $A_0$ and $B_0$ respectively. Any information that is sent from Alice and Bob at center C is accessible to both $A_0$ and $B_0$. All the quantum operations performed at center C are assumed to be performed jointly by $A_0$ and $B_0$. However, no cheating strategies can be applied on flying quantum information from Alice and Bob to center C; $A_0$ and $B_0$ (or authority C) can securely communicate with respective agencies Alice and Bob. Explicit procedure for multiparty commitment is described below:

**Phase-I, t=0:** Both Alice and Bob share randomly chosen EPR pairs $|\alpha\alpha_0\rangle \in \{\zeta,\eta\}$ and $|\beta\beta_0\rangle \in \{\zeta,\eta\}$ with C respectively. By doing this, they commits themselves to 2-bit strings $\alpha\alpha_0$ and $\beta\beta_0$ respectively.

**Phase-II, t=$x$/c:** C performs BSM on qubits in his possession, stores classical BSM result $cc'$ in his memory and broadcast the result as well. This measurement of C projects the qubits in possession of Alice and Bob into Bell states $|\alpha\beta\rangle \in \{\zeta^+, \zeta^-, \eta^+, \eta^-\}$. Simultaneously, Bob teleports a publically known state $|\varphi\rangle \in \{0,1\}$ (either $|\varphi\rangle = |0\rangle$ or $|\varphi\rangle = |1\rangle$) to Alice and gets two classical bits $bb'$. As a result, Alice's half $|\alpha\rangle$ becomes one of the corresponding four possible states $|\psi\rangle = \sigma_i |\varphi\rangle$ where teleportation encoding $\sigma_i \in \{I, \sigma_x, \sigma_z, \sigma_z\sigma_x\}$ is unknown to Alice, Bob and C. Instantly, Alice measures her qubit $|\alpha\rangle$ in {0,1} basis, applies unitary operators $\sigma_z^\alpha \sigma_x^{\alpha_0}$ and returns $|\psi'\rangle = \sigma_z^\alpha \sigma_x^{\alpha_0} \sigma_i |\varphi\rangle$ to center C. Similarly, Bob sends $|\varphi'\rangle = \sigma_z^b \sigma_x^{b'} \sigma_z^\beta \sigma_x^{\beta_0} |\varphi\rangle$ at center C. At this stage, time t=$x$/c, both Alice and Bob confirm their commitments $\alpha\alpha_0$ and $\beta\beta_0$ encoded over $|\psi'\rangle$ and $|\varphi'\rangle$ respectively.

**Phase-III, t=2$x$/c:** C measures $|\psi'\rangle$ and $|\varphi'\rangle$ in the {0,1} basis and stores outcome $\psi'$ and $\varphi'$.

**Phase-IV, t=T:** At some agreed time t=T, both Alice and Bob reveal their commitments by announcing identities of $\alpha\alpha_0$ and $\beta\beta_0$ (and $bb'$) respectively. Now C can find the exact identity of swapped entangled state $|\alpha\beta\rangle$, and (hence) the exact teleportation encoding $\sigma_i$. If announcements of Alice and Bob are consistent with non-local correlations, C validates their commitments otherwise aborts.

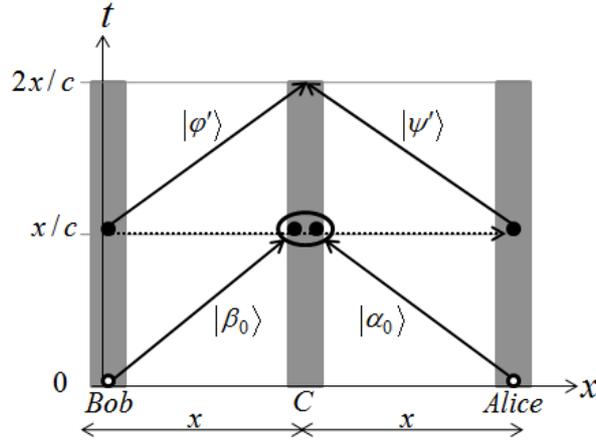

**Figure 2:** Relativistic quantum multiparty bit commitment scheme between Alice and Bob. C represents a joint center where both Alice and Bob have their representatives $A_0$ and $B_0$ respectively. At priorly agreed time t=0, both Alice and Bob make commitments by sharing $|\alpha_0\rangle$ and $|\beta_0\rangle$ with C. Both Alice and Bob confirm their commitment by sending $|\psi'\rangle = \sigma_z^\alpha \sigma_x^{\alpha_0} \sigma_i |\varphi\rangle$ and $|\varphi'\rangle = \sigma_z^b \sigma_x^{b'} \sigma_z^\beta \sigma_x^{\beta_0} |\varphi\rangle$ to C at some later time t=x/c. Solid arrows represent flying qubits while dotted arrow shows teleportation.

**Security Analysis**

Our scheme is purely relativistic quantum mechanical; it does not require any classical communication between committer and receiver and hence is secure from classical attacks by definition. As for as quantum attacks are concerned, quantum non-local correlations guarantee that the scheme is perfectly concealed and receiver cannot predict or extract committed bit before the opening from committer. Similarly, it guarantees receiver that committer cannot change her committed bit after his/her confirmation since there comes a unique Bell state after Bell measurement over initially shared committed Bell states and hence unique Pauli encoding from teleportation over swapped Bell state.

Instead of giving detailed security proof, we argue here that our proposed scheme is secure purely on the basis of rules of relativistic quantum theory. Repetitive measurements at the secure laboratories of Alice, Bob and their agents (center C) ensure security against Mayers and Lo-Chau quantum attacks[4-7] attacks. Repetitive measurements in priorly agreed basis ensure that quantum states $|\varphi\rangle, |\varphi'\rangle$, and $|\psi'\rangle$ are no more entangled and hence cheating party cannot delay (restricted by no-signaling and measurement of time lapse) or alter his/her commitment by applying unitary transformations. Furthermore, measurements in agreed basis allow both parties to know each other's commitment in the revealing phase with surety; authority C cannot announce biased outcome since both parties have stored non-locally correlated measurement outcomes.

Moreover, in multiparty commitment scheme, we assume that both committer and receiver commit and reveal their commitment simultaneously. This element of simultaneity does not compromise security since flying qubits from Alice and Bob to center C are perfectly encrypted through quantum one-time pad[28,29]. Hence neither party can extract any useful information, and hence cannot change his/her commitment accordingly, about committed bit of other party.

Since, $|\varphi\rangle=|0\rangle$ and $|\varphi\rangle=|1\rangle$ are eigenstates of both Pauli operators $I$ and $\sigma_z$, hence equations (2.1), (2.3), (3.1) and (3.3) show that committer can change his/her committed 2-bit string from $d_1d_2=00$ to $d_1d_2=10$ by announcing initially prepared Bell state $\zeta^-$ instead of $\zeta^+$ or vice versa. Since $|\psi'\rangle=I\sigma_i|\varphi\rangle=\sigma_z\sigma_i|\varphi\rangle$, receiver cannot detect cheating committer if $|\varphi\rangle=|0\rangle$ or $|\varphi\rangle=|1\rangle$. However it will not be considered as successful cheating since both Bell state $\zeta^-$ and $\zeta^+$ or $d_1d_2=00$ and $d_1d_2=10$ represents same committed bit $d_c=0$ according to code. Similarly, equations (2.2), (2.4), (3.2) and (3.4) show that committer can announce $\eta^-$ instead of $\eta^+$ or vice versa where both Bell states $\eta^-$ and $\eta^+$ represent same committed bit $d_c=1$ and $|\psi'\rangle=\sigma_x\sigma_i|\varphi\rangle=\sigma_z\sigma_x\sigma_i|\varphi\rangle$ if $|\varphi\rangle=|0\rangle$ or $|\varphi\rangle=|1\rangle$. However, from equations (2.1)-(2.4) and (3.1)-(3.4), it can be seen that committer cannot change his/her initially prepared Bell state from $\zeta^\pm$ to $\eta^\pm$. Hence he/she cannot change his/her commitment from $d_1d_2=\{00,10\}$ to $d_1d_2=\{01,11\}$ or committed bit from $d_c=0$ to $d_c=1$.

Finally, what if receiver Bob and his agent $B_0$ perform not a single designated action in the first round of single-party bit commitment scheme and wait for second round to extract committed bit from Alice? That is neither $B_0$ performs Bell measurement on $|\alpha_0\rangle$ and $|\beta_0\rangle$ nor Bob teleports state $|\varphi\rangle$ to Alice. The confirmation of commitment through unitary transformations $\sigma_z^\alpha \sigma_x^{\alpha_0}$ on her retained half $|\alpha\rangle$ guarantees Alice that $B_0$ cannot extract exact identity of $|\alpha_0\alpha\rangle$ through Bell measurement on $|\alpha_0\rangle$ and $\sigma_z^\alpha \sigma_x^{\alpha_0}|\alpha\rangle$.

**Commitments of classical bit strings:** Instead of committing single-bit $d_c=0$ to $d_c=1$, Alice can commit classical 2-bit strings $d_c^1d_c^2=00,01,10$ or $11$ by sharing an EPR pair $\zeta^+=|00\rangle$, $\eta^+=|01\rangle$, $\zeta^-=|10\rangle$ or $\eta^-=|11\rangle$ with Bob's agent $B_0$ respectively. The single-party scheme can be modified for commitment of any one of classical 2-bit strings $d_c^1d_c^2=\{00,01,10,11\}$ with following code:

**Code:** Suppose Bob and his agent $B_0$ secretly decide quantum state $|\varphi\rangle$ prepared in measurement basis $\{0,1,+,-\}$ somewhere in their causal past unknown to committer Alice. To commit classical 2-bit string 00, 01, 10, or 11, Alice shares an EPR pair $\zeta^+=|00\rangle, \eta^+=|01\rangle, \zeta^-=|10\rangle$ or $\eta^-=|11\rangle$ with Bob's agent $B_0$ respectively.

If Alice wants to commit one of four classical 2-bit strings 00,01,10,11, Bob teleports $|\varphi\rangle\in\{0,1,+,-\}$ to Alice at time $t=x/c$. Since Alice does not know whether $|\varphi\rangle\in\{0,1\}$ (eigenstates of $I$ and $\sigma_z$) or $|\varphi\rangle\in\{+,-\}$ (eigenstates $\sigma_x$ and $\sigma_x\sigma_z$), she cannot measure/store her retained entangled half now in the state $|\psi\rangle=\sigma_i|\varphi\rangle$. If Alice tries to cheat by announcing Bell state $|\alpha'\alpha_0'\rangle$ different from initially shared $|\alpha\alpha_0\rangle$, her success probability will only be ½ since there are only 50% chance that $\sigma_z^{\alpha'}\sigma_x^{\alpha_0'}\sigma_i|\varphi\rangle=\sigma_z^\alpha \sigma_x^{\alpha_0}\sigma_i|\varphi\rangle$. If N entangled states are used, Alice can commit for

2N classical bit string $|\alpha\alpha_0\rangle_1 |\alpha\alpha_0\rangle_2 .........|\alpha\alpha_0\rangle_N$ where Alice's cheating probability asymptotically goes to zero; $1/2^N \to 0$ for large N.

**Applications**
Outcomes of the proposed relativistic quantum commitment scheme such as indefinite commitment time, commitment of arbitrarily long bit strings, and allowing multiparty commitments would have a number of applications in business and secure multiparty computation as follows.

**Joint bank accounts:** Alice and Bob can be considered as two users of joint account at bank C. Both Alice and Bob commit with each other and with bank C by sharing their Bell pairs and later confirming their willing (committed bit $d_c$) to bank. In view of commitments from both account holders, bank C can release or halt transactions to anyone of these parties or both.

**Stock exchange:** Alice and Bob can be treated as two business companies who want to invest jointly for certain shares in stock exchange but each have interest in keeping his/her share secret from other. They can commit at stock exchange simultaneously and later open their commitments without distressing each other's interests. They cannot enhance or reduce their shares by looking their interests in current market trends but can take decisions only that are allowed by the protocol; they can invest either one of $d_1 d_2 = 00$ and $d_1 d_2 = 10$ if initially committed to bit $d_c = 0$ or one of $d_1 d_2 = 01$ and $d_1 d_2 = 11$ if initially committed to bit $d_c = 1$.

**Blind bidding:** Suppose authority C announces bidding for some of his articles. Alice and Bob are two distant competitors (bidders) who are asked to blindly bid for those articles without knowing each other's bid. They secretly share their concealed bids (commitments) by sending entangled halves of their Bell states $|\alpha\alpha_0\rangle \in \{\zeta, \eta\}$ and $|\beta\beta_0\rangle \in \{\zeta, \eta\}$ to C respectively. At the end of the protocol, C announces which party has won the bidding by publically announcing final outcome that comes from pre-defined rule. Here Alice or Bob can challenge the authority C if he/she founds that announcement of winner was biased.

**Multiparty secure computation:** More generally, proposed scheme can be used for multiparty secure computation without assuming any bound on the corrupt competitors (Alice and Bob) or authority (C). At the end of the scheme, input from each party remains secret while everyone knows a definite outcome. That is, the authority C can validate or abort the committed bit $d_c = 0$ to $d_c = 1$ and announce the outcome but cannot extract whether $d_c$ has come from $d_1 d_2 = \{00, 10\}$ or $d_1 d_2 = \{01, 11\}$.

**Discussion**
We proposed a delayed choice relativistic bit commitment scheme that enhances commitment time to be arbitrarily long and can be used for multiparty commitment where both parties commit and reveal simultaneously. The same commitment scheme can also be used for the commitment of arbitrarily long classical bit strings. The unconditional security from both committer and receiver comes through non-local quantum correlation generated by very fundamental and important quantum information tasks, entanglement swapping and teleportation.

In the previously proposed relativistic bit commitment schemes, authors tried to enhance commitment time either through multiple rounds of communications or by enhancing communication time by taking large distance between committer and receiver. However, our proposed relativistic quantum bit commitment scheme gives arbitrarily long commitment time by using only two round of communication while distance *x* between committer and receiver can be as small as possible.

The proposed relativistic scheme is practical and can be efficiently applied for arbitrarily long commitment time with existing technologies without requiring long term quantum memories or maintaining coherence over distant quantum channels. Repetitive measurements allow to store measurement outcomes that are non-locally correlated with initially prepared entangled states; entanglement is required only for time $t=x/c$ where *x* is the special distance between the committer and receiver.

Hopefully, practically relevant outcomes from this scheme will shed light on relationship between quantum information theory, especially quantum cryptography, with relativistic quantum theory.